\documentclass[sigconf,nonacm]{acmart}

\usepackage{multirow}
\usepackage[linesnumbered,ruled,vlined]{algorithm2e}
\usepackage{flushend}
\begin{document}
\title{RemoteANN: A Native Vector Search Engine for Cloud Object Storage}

\author{Yutong Gou}
\affiliation{\institution{Nanyang Technological University}\country{Singapore}}
\email{yutong003@e.ntu.edu.sg}

\author{Silu Huang}
\affiliation{\institution{ByteDance}\country{USA}}
\email{silu.huang@bytedance.com}

\author{Yue Wang}
\affiliation{\institution{ByteDance}\country{USA}}
\email{yue.w@bytedance.com}

\author{Jianjun Chen}
\affiliation{\institution{ByteDance}\country{USA}}
\email{jianjun.chen@bytedance.com}

\author{Cheng Long}
\affiliation{\institution{Nanyang Technological University}\country{Singapore}}
\email{c.long@ntu.edu.sg}

\begin{abstract}
Vector search engines built on cloud object storage have drawn growing attention from both academia and industry.
These systems typically adopt a disaggregated storage-compute architecture, persisting the primary vector payload in object stores (e.g., Amazon S3) while retaining compact index metadata on compute nodes.
However, conventional vector indexing structures are not natively tailored to the pronounced network latency and per-request access costs of cloud object stores, resulting in network I/O amplification, degraded query throughput, and higher API expenses.

To resolve these challenges, we present \textit{RemoteANN}, a vector search engine explicitly co-designed for cloud object storage across indexing and querying.
To mitigate the dominant network I/O bottleneck, RemoteANN couples an Inverted File (IVF) index with RaBitQ quantization, further optimized by inter-cluster and intra-cluster data reordering for improved data locality.
Building on this tailored data layout, RemoteANN's query engine combines cluster pruning, fine-grained Range-GET requests, request merging, and local NVMe SSD caching, dramatically reducing discrete storage requests and network transfer volume without sacrificing accuracy.
To address the index construction bottleneck, we further develop \textit{QGKMeans}, which substantially reduces construction latency without compromising clustering quality.
Extensive evaluations on real-world datasets demonstrate that, compared with representative baselines, RemoteANN achieves up to a $5.9\times$ throughput speedup while issuing up to $16.1\times$ fewer object storage requests at 97\% target recall.
\end{abstract}

\maketitle
\pagestyle{plain}
\thispagestyle{plain}
\raggedbottom

\section{Introduction}
\label{sec:intro}

Cloud object storage platforms (e.g., Amazon S3~\cite{awss3}, Azure Blob Storage~\cite{azureblob}, and Torch Object Storage~\cite{tos}) have become an attractive storage backend for modern database systems, owing to their horizontal scalability, high data durability, and cost efficiency.
By decoupling compute from storage, this architecture allows stateless compute instances (e.g., AWS EC2~\cite{awsec2}) to handle query execution while offloading primary data to elastic, remote object stores.
Motivated by these infrastructure benefits, both academia and industry have sought to extend this paradigm to vector database systems and approximate nearest neighbor (ANN) search engines (e.g., Amazon S3 Vectors~\cite{s3vec}, Turbopuffer~\cite{turbopuffer}, Lance~\cite{lancedb}, Databricks~\cite{databricks}).
In these systems, compact index metadata and routing structures reside in the compute nodes' local storage, while the bulk of high-dimensional vector payloads is persisted in cloud object storage.

However, conventional vector index structures are not natively suited to cloud object storage, and deploying them directly leads to suboptimal query performance and elevated operational cost.
First, compared to local storage media (e.g., DRAM and local SSDs), data retrieval from cloud object stores incurs substantial time-to-first-byte (TTFB) network latency, which dominates total download latency for small payload requests~\cite{anyblob}.
As a result, graph-based index structures~\cite{hnsw,nsg,2019diskann} suffer severe performance degradation due to their iterative, multi-hop routing patterns, 
and existing vector search systems on cloud object storage predominantly adopt cluster-based indices instead (e.g., IVF in Lance~\cite{lancedb} and SPFresh~\cite{spfresh} in turbopuffer~\cite{turbopuffer}).
Second, cloud providers charge substantially for the number of discrete storage requests issued, which is a cost dimension that traditional, locally-oriented index designs do not account for.
A naive implementation can therefore trigger excessive request volumes, leading to prohibitive API costs.
Despite this, no existing system jointly optimizes index layout, index construction, and query execution for the latency and cost characteristics of cloud object storage.

To close this gap, we present RemoteANN, a vector search engine explicitly designed for cloud object storage.
Through an algorithm-system co-design spanning indexing, querying, and the storage hierarchy, 
RemoteANN systematically mitigates the network I/O bottlenecks that limit vector search on cloud object storage.
Specifically, RemoteANN makes the following contributions in both indexing and querying:

\begin{itemize}
    \item \textbf{Query-Oriented Indexing and Data Layout.}\par 
    RemoteANN combines an IVF index with RaBitQ quantization~\cite{rabitq,exrabitq}.
    Following existing frameworks, it keeps index metadata and 1-bit RaBitQ codes in DRAM while storing primary vector payloads in cloud object storage.
    The memory-resident codes support pruning before remote data retrieval, while inter-cluster and intra-cluster data reordering improve physical locality.
    These changes substantially improve the effectiveness of our I/O optimization strategies during querying.
    To reduce the cost of indexing, we develop QGKMeans, which uses SymphonyQG~\cite{gou2025symphonyqg} to accelerate centroid assignment during $k$-means clustering.
    This substantially reduces index construction latency without compromising clustering quality.

    \item \textbf{Co-Designed Query Engine.}\par 
    RemoteANN incorporates a multi-tier query optimization framework tailored to cloud object storage, including RaBitQ-driven cluster pruning that prunes clusters based on
    analytical error bounds provided by RaBitQ, range-optimized Range-GET requests that download only the minimal candidate byte slices, request merging to reduce API cost, and a local NVMe SSD caching layer that bypasses remote I/O. 
    Built upon our co-designed data layout, these components jointly reduce cloud I/O overhead without sacrificing query performance.
\end{itemize}

We evaluate RemoteANN on diverse real-world datasets spanning a range of dimensions and scales.
The results confirm the effectiveness of our environment-specific design for both indexing and querying.
In extensive experiments on real-world datasets, RemoteANN achieves up to a $5.9\times$ throughput speedup while issuing up to $16.1\times$ fewer discrete storage requests compared with representative baselines.
\section{Preliminary}
\label{sec:pre}

\begin{figure}
  \centering
  \includegraphics[width=\linewidth]{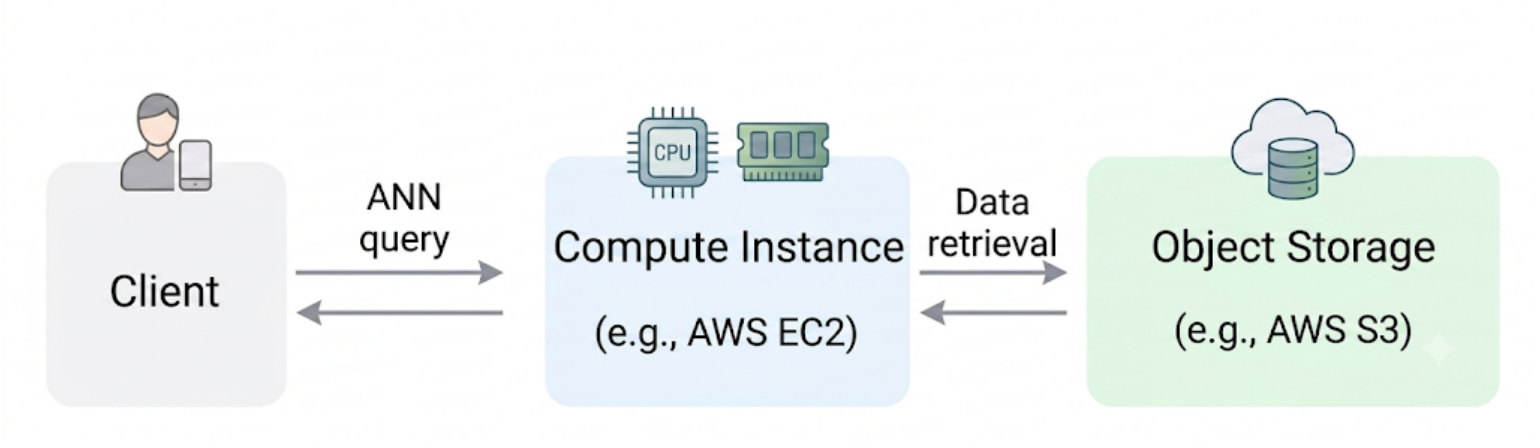}
  \Description{Vector search on cloud object storage.}
  \caption{Vector search on cloud object storage.}
  \label{fig:cloudvec}
\vspace{-2mm}
\end{figure}

\subsection{Cloud Object Storage}
\label{subsec:cloud}
Cloud object storage refers to the online object storage service offered by cloud vendors~\cite{awss3,azureblob,tencentobj,tos,gcs,oracleobj,ibmobj}. 
By orchestrating a distributed environment across diverse storage nodes and servers, 
cloud object storage delivers high data durability, seamless horizontal scalability, and exceptional cost-efficiency, rendering it ideal for hosting massive scales of unstructured data.

Cloud object storage differs fundamentally from local storage across two primary aspects: 
the pricing model and underlying I/O characteristics, both of which require careful design of the system built on it. 
First, regarding data retrieval operations, cloud vendors mainly charge the user by the number of requests~\footnote{Here, we assume the data retrieval request originates from instances co-located within the same cloud provider's region. Otherwise, the data transfer cost can significantly exceed the request fees, which makes the system financially non-viable.}.
Second, the retrieval operations incur high latency due to the time-to-first-byte (TTFB) overhead (typically tens of milliseconds) introduced by network communication. 
As established in a previous study~\cite{anyblob}, this TTFB penalty dominates total latency for small payload sizes.

\subsection{Vector Search on Cloud Object Storage}
\label{subsec:cloudnative}

The inherent advantages of cloud object storage have driven interest from both academia and industry in building ANN search engines on it (e.g., Amazon S3 Vectors~\cite{s3vec}, Turbopuffer~\cite{turbopuffer}). 
As illustrated in Figure~\ref{fig:cloudvec}, 
these systems typically contain two core layers: (i) a compute instance utilizing local storage (memory and disk) to cache index metadata, 
and (ii) a cloud object store responsible for persisting the primary index payload data. 
To mitigate the pronounced TTFB latency overhead intrinsic to cloud object stores, these engines predominantly adopt cluster-based ANN indices (such as IVF variants and SPFresh~\cite{douze2024faiss,spfresh,s3vec,turbopuffer}) 
rather than popular graph-based indices~\cite{hnsw,nsg,2019diskann} for in-memory or disk-based scenarios.
This is because cluster-based indices can fetch all target vector data within a single network round-trip from the cloud object storage. 
Conversely, the iterative routing pattern of graph-based indices necessitates multi-round-trip network traversals, leading to cumulative TTFB overhead and inflated end-to-end query latency.

\subsection{RaBitQ Quantization}
\label{subsec:rabitq}
For vector search, quantization constitutes a class of techniques designed to compress high-dimensional vectors into compact, 
low-bit quantization codes~\cite{pq,pqfs,opq,addq,guo2020accelerating,babenko2014inverted,rabitq,exrabitq,turboquant,drive,eden,mrq}. 
During query processing, the system uses these compressed codes for distance estimation.
Recently, a novel quantization framework named RaBitQ~\cite{rabitq,exrabitq,rabitqlib} was proposed to address the limitation (i.e., the lack of theoretical support) of classical quantization methods. 

During indexing for $D$-dimensional vectors, 
RaBitQ constructs a codebook with $2^{B \cdot D}$ unit vectors using a uniform grid, 
where $B$ denotes the number of bits allocated per dimension.
Subsequently, the codebook is randomly rotated via Johnson-Lindenstrauss Transformation~\cite{johnson1984extensions},
and each data vector is normalized and then quantized to its nearest codeword within the codebook.
For querying, 
RaBitQ estimates the distances between a query vector and data vectors based on their codewords.
Based on rigorous theoretical analysis,
the RaBitQ estimator provides an unbiased estimation with a theoretical error bound,
enabling superior empirical retrieval accuracy across real-world workloads~\cite{exrabitq}.

\section{Methodology}
\label{sec:method}

In this section, we present the comprehensive system design of our cloud vector search framework called RemoteANN, 
discussing both index construction and query processing optimizations tailored for the specific environment. 
Due to the high I/O latency of the cloud object storage discussed in Section~\ref{subsec:cloud},
we choose the IVF index since it enables single-round-trip target cluster retrieval for a given query.
To mitigate the remaining cloud I/O overheads, we introduce coordinated architectural optimizations spanning both indexing and querying.

We present details of our indexing pipeline in Section~\ref{subsec:index}, 
including the heterogeneous storage layout configured with RaBitQ quantization (Section~\ref{subsubsec:layout}), 
our efficient IVF clustering acceleration (Section~\ref{subsubsec:symqg})
and data reordering strategies to boost data locality and support query optimization strategies (Section~\ref{subsubsec:reorder}). 
Section~\ref{subsec:query} then details our proposed query processing framework. 
Specifically, for an IVF index deployed on cloud object storage, the query engine initially identifies the $n_\text{probe}$ target clusters for a given query vector via the metadata resident on the compute node, 
subsequently fetching the corresponding data payloads from cloud object storage. 
To reduce cloud I/O overhead, our engine incorporates four optimization strategies: namely RaBitQ-based cluster-level pruning (Section~\ref{subsubsec:prune}), fine-grained 
range optimization (Section~\ref{subsubsec:range}), request merging (Section~\ref{subsubsec:merge}), and caching (Section~\ref{subsubsec:cache}).
These multi-tier components are structurally co-designed; the layout properties established during index generation directly dictate and unlock the high-efficiency I/O primitives leveraged throughout query processing.
Section~\ref{subsec:sum} provides a summary of the indexing and querying components of RemoteANN.

\subsection{The Index Phase}
\label{subsec:index}

\subsubsection{Storage Layout of IVF index with RaBitQ Quantization}
\label{subsubsec:layout}
As established in Section~\ref{sec:pre}, vector search on cloud object storage contains two components -- 
the compute node and the cloud object storage, thereby establishing a distinct three-layer storage hierarchy 
consisting of local DRAM, local disk, and remote storage.
To optimize operational costs, the system may persist all data exclusively within cloud object 
storage during periods of inactivity.
When a query is initiated, the system needs to spin up a compute instance and pull required data from the cloud
object storage through the network,
which results in a severe cold-start penalty with prohibitive initial query latency. 
We note that this latency can be mitigated via strategic system-level co-design. 
For instance, when integrating vector search engines with interactive downstream applications 
(e.g., conversational AI agents), the user's interaction (e.g., typing) can serve as early pre-fetching signals. 
The backend system captures these initial signals to asynchronously initiate the download of index metadata, 
effectively hiding the cold-start latency.
Therefore, in this study, we restrict our focus to the warm query processing paradigm, 
where the high-precision vector data resides permanently within the cloud object storage, whereas frequently accessed index metadata 
(e.g., the IVF cluster centroids) are pinned within the compute node's main memory during query processing.
We detail our data layout framework as follows. 

The compute node's DRAM caches the high-dimensional centroid vectors and the low-precision quantized data vectors as a routing index. 
Specifically, we leverage the RaBitQ quantization~\cite{rabitq, exrabitq} to compress high-dimensional 
raw vectors into a 9-bit representation\footnote{We configure a 9-bit quantization because empirical results reported 
in the paper~\cite{exrabitq} demonstrate that the 9-bit code successfully achieves similar retrieval accuracy 
as the uncompressed vectors across diverse datasets. 
Depending on the specific workload, users can reduce the bit allocation.}.
Furthermore, we partition this 9-bit code into a 1-bit prefix and an 8-bit suffix. 
The 1-bit prefixes are resident in local DRAM, whereas the corresponding 8-bit suffixes are persisted in cloud object storage. 
If the compute instance is configured with local NVMe SSD, the system can utilize it as an intermediate cache 
for the 8-bit suffixes of frequently accessed clusters.
The primary objective of this decoupled quantization scheme is to optimize cloud object storage I/O 
 -- the dominant performance bottleneck of the whole system. 
The RaBitQ framework delivers high-accuracy distance estimation with rigorous theoretical guarantees. 
Consequently, the query engine can exploit the in-memory 1-bit prefixes to prune vectors and clusters, 
thereby reducing both the total number of distinct Range-GET requests issued to cloud object storage 
and the data volume downloaded (detailed in Section~\ref{subsec:query}).

\begin{table}[tbp]
\setlength{\tabcolsep}{2.5pt}
\footnotesize
\centering
\caption{Evaluation of different $k$-means variants. The choice of $k$ follows the recommendation in Faiss~\cite{douze2024faiss}. 
Downstream effectiveness is quantified via top-10 ANN search recall on the IVF index built from the generated clusters, evaluating a fixed 1\% cluster inspection fraction.
Results for baselines are omitted on Deep-1B since all of them exceeded the 12-hour time threshold.
}
\label{tab:kmeans}
  \vspace{-2mm}
\resizebox{1.0\columnwidth}{!}{\begin{tabular}{l l c c c c}
\toprule
\textbf{Dataset}&
\textbf{Method} &
\textbf{$k$} &
\textbf{Time (s)} &
\textbf{Recall} &
\textbf{WCSS} \\
\midrule
\multirow{4}{*}{SIFT}
 & QGKMeans   & \multirow{4}{*}{4096} & \textbf{6.2} & 0.941 & $4.442 \times 10^{10}$ \\
 & Faiss      &                       & 13.0         & 0.940 & $4.440 \times 10^{10}$ \\
 & SuperKMeans&                       & 8.9          & 0.941 & $4.442 \times 10^{10}$ \\
 & SeededANN  &                       & 19.0         & 0.940 & $4.442 \times 10^{10}$ \\
\midrule
\multirow{4}{*}{MSong}
 & QGKMeans   & \multirow{4}{*}{4096} & \textbf{9.2} & 0.992 & $5.086 \times 10^{8}$ \\
 & Faiss      &                       & 38.8         & 0.994 & $5.094 \times 10^{8}$ \\
 & SuperKMeans&                       & 10.1         & 0.991 & $5.092 \times 10^{8}$ \\
 & SeededANN  &                       & 33.6         & 0.992 & $5.096 \times 10^{8}$ \\
\midrule
\multirow{4}{*}{Contriever}
 & QGKMeans   & \multirow{4}{*}{4096} & \textbf{12.9} & 0.977 & $6.281 \times 10^{5}$ \\
 & Faiss      &                       & 63.5          & 0.978 & $6.279 \times 10^{5}$ \\
 & SuperKMeans&                       & 28.2          & 0.978 & $6.280 \times 10^{5}$ \\
 & SeededANN  &                       & 75.3          & 0.978 & $6.280 \times 10^{5}$ \\
\midrule
\multirow{4}{*}{GIST}
 & QGKMeans   & \multirow{4}{*}{4096} & \textbf{15.6} & 0.821 & $1.035 \times 10^{6}$ \\
 & Faiss      &                       & 80.7          & 0.816 & $1.035 \times 10^{6}$ \\
 & SuperKMeans&                       & 23.2          & 0.822 & $1.035 \times 10^{6}$ \\
 & SeededANN  &                       & 94.9          & 0.818 & $1.036 \times 10^{6}$ \\
\midrule
\multirow{4}{*}{MXBAI}
 & QGKMeans   & \multirow{4}{*}{4096} & \textbf{14.0} & 0.967 & $9.938 \times 10^{7}$ \\
 & Faiss      &                       & 68.2          & 0.967 & $9.938 \times 10^{7}$ \\
 & SuperKMeans&                       & 16.4          & 0.965 & $9.937 \times 10^{7}$ \\
 & SeededANN  &                       & 66.6          & 0.967 & $9.938 \times 10^{7}$ \\
\midrule
\multirow{4}{*}{OpenAI}
 & QGKMeans   & \multirow{4}{*}{4096} & \textbf{24.6} & 0.895 & $6.205 \times 10^{5}$ \\
 & Faiss      &                       & 130.4         & 0.896 & $6.204 \times 10^{5}$ \\
 & SuperKMeans&                       & 29.4          & 0.897 & $6.203 \times 10^{5}$ \\
 & SeededANN  &                       & 142.5         & 0.896 & $6.203 \times 10^{5}$ \\
\midrule
\multirow{4}{*}{Cohere}
 & QGKMeans   & \multirow{4}{*}{65536} & \textbf{204.2} & 0.947 & $4.297 \times 10^{6}$ \\
 & Faiss      &                        & 13144.8        & 0.953 & $4.292 \times 10^{6}$ \\
 & SuperKMeans&                        & 2330.3         & 0.951 & $4.293 \times 10^{6}$ \\
 & SeededANN  &                        & 1128.4         & 0.951 & $4.293 \times 10^{6}$ \\
\midrule
\multirow{4}{*}{Deep-1B}
 & QGKMeans   & \multirow{4}{*}{1048576} & \textbf{8973.5} & 0.999 & $1.678 \times 10^{7}$ \\
 & Faiss      &                          & --              & --    & -- \\
 & SuperKMeans&                          & --              & --    & -- \\
 & SeededANN  &                          & --              & --    & -- \\
\bottomrule
\end{tabular}
}
\vspace{-2mm}
\end{table}

{

\subsubsection{IVF Indexing Acceleration with QGKMeans}
\label{subsubsec:symqg}
Conventionally, constructing the IVF index for ANN search relies on partitioning the dataset via the $k$-means clustering algorithm. 
However, as reported in established studies~\cite{douze2024faiss, annbench, bigann,exrabitq}, 
repeated vector-to-centroid assignments make $k$-means clustering a bottleneck in IVF index construction.
To accelerate the indexing of IVF, prior works have applied techniques from ANN search to reduce the cost of centroid assignment in the $k$-means iterations~\cite{superkmeans,lindormvector,scalakmeans,deng2018knn}.

Inspired by existing studies, we propose a new $k$-means variant named QGKMeans which adopts SymphonyQG~\cite{gou2025symphonyqg} as the accelerator for the expensive vector-to-centroid assignment.
At each $k$-means iteration, we construct a SymphonyQG index over the current centroids and query it with each data vector to find its approximate nearest centroid.
This setting requires many searches over each $k$-means iteration, followed by rebuilding the search structure when the centroids are updated.
As reported in previous studies and benchmarks~\cite{gou2025symphonyqg,vibebench}, SymphonyQG has demonstrated very competitive efficiency for both querying and indexing, 
which makes it a good candidate for the accelerator.

We compare QGKMeans with Faiss~\cite{douze2024faiss}, SuperKMeans~\cite{superkmeans}, and SeededANN~\cite{scalakmeans} in terms of clustering time and the quality of the resulting clusters.
Faiss provides a widely used reference implementation.
SuperKMeans targets the same workload through dimension pruning~\cite{adsampling} and an optimized data layout~\cite{kuffo2025pdx}, 
while SeededANN represents an HNSW-based approach to accelerating centroid assignment.
All variants use the full set of original vectors and run for 25 iterations. 
We disable data sampling and hierarchical clustering (as they are easy to combine with all $k$-means variants) to compare them under the same flat $k$-means workload. 
Following the evaluation criteria in~\cite{superkmeans}, we measure total clustering time, Within-Cluster Sum of Squares (WCSS), and downstream IVF recall when probing a fixed fraction of clusters.

All $k$-means variants are evaluated on an Intel(R) Xeon(R) Gold 6418H CPU, pinning the execution workload to 48 concurrent threads. 
The empirical results are summarized in Table~\ref{tab:kmeans}. 
As illustrated, QGKMeans achieves the lowest clustering time among the evaluated methods while maintaining comparable WCSS and downstream recall. 
Its speedup over the fastest baseline is $1.4\times$ on SIFT and $5.5\times$ on Cohere. 
We therefore use QGKMeans to generate clusters throughout all subsequent experiments.}

\subsubsection{Data Reordering for Facilitating Query Processing}
\label{subsubsec:reorder}
{ In conventional in-memory or disk-based IVF storage layouts, 
multiple clusters share a large binary file, and the query engine retrieves the selected clusters from this file.
Both the order of clusters within the file and the order of vectors within each cluster are typically arbitrary,
as efficient random access in RAM and SSDs makes these layouts less sensitive to physical ordering.
However, directly adopting this layout and access pattern in cloud object storage introduces two issues.}

First, because the vendors charge users of cloud object storage by the number of requests, issuing discrete Range-GET requests
for every target cluster becomes financially non-viable, particularly for high-recall queries that necessitate traversing many clusters.
A straightforward solution is to consolidate multiple target clusters into a unified request, thereby substantially reducing the query cost.
However, if the target clusters for a given query exhibit poor spatial locality (i.e., they are sparsely distributed across the index object), 
request consolidation becomes highly inefficient, as the growing request size amplifies the overall download latency.
Second, monolithic downloading for every target cluster introduces severe I/O inefficiencies. 
An individual IVF cluster may contain hundreds or even thousands of vectors, 
yet a given query typically finds only a small fraction of them as promising nearest neighbor candidates. 
Consequently, always downloading the whole cluster results in a waste of network bandwidth and degraded throughput, especially for the workload with many concurrent queries.

To address these two problems, we introduce two query-side optimization strategies: \textit{request merging} and \textit{range optimization}. 
The former consolidates disparate cluster requests into a unified one, while the latter reduces the data volume transferred per cluster. 
We detail these query processing mechanisms in Sections~\ref{subsubsec:merge} and~\ref{subsubsec:range}, respectively; 
this section focuses on the indexing part required to support them. 
Specifically, we examine how to reorder data at the inter-cluster and intra-cluster levels to boost our query optimizations.

\subsubsection*{\textbf{Inter-cluster reordering}}
{ To enable effective request merging, we need to reorder clusters within the IVF object so that clusters likely to be accessed together are placed close to one another.
Recall that IVF selects target clusters based on the distances between the query vector and the cluster centroids.
Intuitively, clusters with nearby centroids are more likely to be selected together by a query, 
motivating us to place them close to one another in the IVF object.

Assuming no prior knowledge of the query workload, we determine the cluster order solely from the centroids.
Specifically, we seek an ordering that minimizes the sum of distances between the centroids of consecutive clusters.
We formulate this objective as a minimum-weight Hamiltonian path problem, where each vertex represents a cluster and each edge
is weighted by the Euclidean distance between the corresponding centroids.
We then seek an approximate minimum Hamiltonian path under the Euclidean distance metric. }
Because the minimum Hamiltonian path problem is NP-hard, exact solutions are impractical for large cluster counts 
(the best-known classical exact algorithm based on dynamic programming runs in $O(n^2 2^n)$ time~\cite{garey2002computers}, where $n$ is the number of vertices/clusters).
Given that the edge weights satisfy the triangle inequality, 
we can leverage polynomial-time approximation frameworks to achieve high-quality results. 
Specifically, we adopt two foundational algorithms: 
the nearest-neighbor heuristic, which provides an $O(\log n)$ worst-case approximation~\cite{rosenkrantz1977analysis}, and the MST-doubling algorithm, which provides a 2-approximation~\cite{hoogeveen1991analysis}. 
Both algorithms require an initial distance matrix computation, leading to a total time complexity of $O(n^2d)$ for $d$-dimensional vectors. 
While more advanced algorithms achieve tighter theoretical bounds (e.g.,~\cite{an2015improving,zenklusen20191}), 
they introduce substantial computational overhead, which degrades the efficiency for index construction. 
Additionally, we evaluate space-filling curves,
specifically the Hilbert and Z-order curves~\cite{sagan2012space, hilbert, zorder},
as alternative baselines because of their established utility for data ordering and nearest-neighbor search~\cite{chen2005neighbor, liao2001high}.
As space-filling curves suffer from the curse of dimensionality, 
we introduce a dimensionality reduction by Principal Component Analysis (PCA) to project the high-dimensional cluster centroids into a 3-dimensional subspace before using them for reordering. 

\begin{table}[t]
  \centering
  \caption{Evaluation of cluster reordering strategies. The Hamiltonian path length is normalized by the baseline.}
  \label{tab:reorder}
  \vspace{-2mm}
  \small
  \setlength{\tabcolsep}{4pt}
  \begin{tabular}{lcccc}
    \toprule
    & \multicolumn{4}{c}{\textbf{Hamiltonian Path Length (normalized)}} \\
    \cmidrule(lr){2-5}
    \textbf{Dataset} &  \shortstack{\textbf{Nearest-}\\\textbf{neighbor}} & \textbf{MST} & \textbf{Hilbert} & \textbf{Z-order} \\
    \midrule
    SIFT        & \textbf{0.3667}  & 0.4157 & 0.5733 & 0.5798  \\
    MSong       & \textbf{0.3583}  & 0.4005 & 0.5567 & 0.5693  \\
    Contriever  & \textbf{0.6413}  & 0.6739 & 0.8796 & 0.8824  \\
    GIST        & \textbf{0.4636}  & 0.5261 & 0.6470 & 0.6560  \\
    MXBAI       & \textbf{0.5476}  & 0.5812 & 0.8152 & 0.8198  \\
    OpenAI      & \textbf{0.5685}  & 0.5980 & 0.8261 & 0.8332  \\
    \bottomrule
  \end{tabular}
\vspace{-2mm}
\end{table}

We evaluate the four aforementioned cluster reordering strategies on six million-scale datasets. 
We utilize QGKMeans to construct the initial IVF index with $4,096$ clusters, where the resulting centroid sequence naturally exhibits an arbitrary, 
randomized baseline order. 
We subsequently execute our four reordering strategies over these centroids. 
To quantify the quality of inter-cluster reordering, we compute the total Hamiltonian path length 
(i.e., the cumulative Euclidean distance across all adjacent pairs in the global cluster sequence)
for both the baseline and optimized configurations. 
The results are presented in Table~\ref{tab:reorder}.
As illustrated, the nearest-neighbor heuristic achieves the minimum Hamiltonian path length across all evaluated datasets, 
despite the fact that the MST-doubling algorithm offers a superior theoretical worst-case approximation. 
Conversely, the space-filling curves exhibit significantly longer Hamiltonian paths. 
Therefore, our framework implements the nearest-neighbor heuristic as the default cluster reordering method.
{ Section~\ref{subsubsec:merge} evaluates how different inter-cluster reordering methods influence request merging during
ANN search, which presents the effectiveness of reordering by reducing the Hamiltonian path length. }

To mitigate the $O(n^2d)$ computational overhead encountered on large-scale datasets with many centroids, we introduce a preliminary centroid partitioning process~\footnote{The approximation guarantee above applies to the unpartitioned heuristic; we use the partitioned variant as a practical optimization for large centroid sets.}.
Specifically, we employ a recursive bisection $k$-means strategy to progressively subdivide the centroids until the cardinality of each leaf node is less than a predefined constraint.
Subsequently, the localized nearest-neighbor heuristic is applied independently within each partition, and the resulting sub-sequences are coalesced according to the hierarchical tree traversal order to finalize the global cluster sequence.
When evaluated on $262,144$ centroids, the whole reordering process completes in less than $1$ minute, demonstrating operational viability for large datasets.

\subsubsection*{\textbf{Intra-cluster reordering}}
{ Range optimization aims to selectively download only the specific byte range within an IVF cluster that contains promising nearest-neighbor candidates for a given query.
To boost the range optimization, we sort vectors within each cluster by their Euclidean distance to the centroid.
Intuitively, when a cluster's centroid is far from the query, vectors near the centroid are less likely to contribute to the final top-$k$ results, 
while peripheral vectors may remain relevant.
This ordering groups peripheral vectors into a contiguous region, potentially reducing the byte range that needs to be downloaded.
Moreover, our RaBitQ-based range optimization also benefits from distance-based intra-cluster reordering,
which will be further explained and evaluated in Section~\ref{subsubsec:range}.}
\subsection{The Query Phase}
\label{subsec:query}

\subsubsection{Cluster Pruning with Coarse Distance Estimator of RaBitQ}
\label{subsubsec:prune}
As established in Section~\ref{subsec:index}, the 1-bit RaBitQ codes are resident within the main memory of the compute node, 
serving as a coarse-grained distance estimator to preclude downloading clusters that contain 
no viable candidates for the final ANN results.
We formalize this process as \textit{cluster pruning} and detail its mechanics below.

By leveraging these 1-bit RaBitQ codes, the system can compute an estimated distance $dist_{est}$ for
all vectors residing within the target clusters. 
Furthermore, because RaBitQ provides an unbiased estimator with a theoretical error bound $err$~\cite{rabitq,exrabitq}, we can derive an analytical upper bound ($dist_{up}$) and lower bound ($dist_{lo}$) for 
each query-to-vector distance:
\begin{equation}
  \begin{aligned}
    dist_{up} = dist_{est} + err,\ \
    dist_{lo} = dist_{est} - err.
  \end{aligned}
\end{equation}
If the true maximum distance of the final top-$k$ ANN results (denoted as $dist_k$) were known \textit{a priori}, 
any vector satisfying $dist_{lo} > dist_k$ could be immediately pruned, as it is non-viable
for such a vector to qualify as one of the $k$ nearest neighbors
~\footnote{Strictly speaking, the error bound provided by RaBitQ~\cite{rabitq,exrabitq} holds with high probability.
Our empirical analysis across multiple datasets confirms that fewer than 0.01\% of vectors violate these analytical bounds per query workload. 
Given the approximate nature of ANN search, we observed no material change in final recall caused by this negligible bound violation rate.}.

Because $dist_k$ cannot be determined prior to retrieving the data from cloud object storage, 
we establish a conservative pruning threshold. 
Specifically, for all candidate clusters, 
we evaluate the upper bounds of all constituent vectors to determine the 
$k$-th smallest
upper distance bound, 
denoted as $\mathit{dist}_k^u$. By definition, $\mathit{dist}_k^u \geq dist_k$. 
Consequently, if a vector's lower bound distance exceeds this threshold ($dist_{lo} > \mathit{dist}_k^u$), 
it will be non-viable under the true top-$k$ distance constraint. 
Extending this logic to cluster-level granularity, if the condition $dist_{lo} > \mathit{dist}_k^u$ holds for all vectors within a given cluster, 
the entire cluster can be pruned, bypassing the corresponding request to cloud object storage entirely. 
Because both $\mathit{dist}_k^u$ and $dist_{lo}$ are evaluated strictly using the RaBitQ codes in DRAM, 
this mechanism effectively reduces request costs and network bandwidth consumption while preserving nearly identical empirical recall.
Alternatively, 
the $k$-th smallest
estimated distance $\mathit{dist}_k^e$ can be employed as an aggressive pruning threshold. 
However, since $\mathit{dist}_k^e$ is not guaranteed to be larger than $dist_k$, 
its application may inadvertently discard valid nearest neighbors, incurring a recall degradation. 
For application scenarios which tolerate this trade-off in accuracy, 
adopting $\mathit{dist}_k^e$ yields higher cluster pruning ratios and further reduces cloud I/O costs.

We evaluate our cluster pruning strategy across six million-scale datasets using their designated query workloads, 
with the results compiled in Table~\ref{tab:prune}. 
As illustrated, the proposed pruning framework significantly reduces the number of clusters to be downloaded, especially for high-dimensional datasets.
For example, on the OpenAI (1536-dimensional) dataset, the query engine retrieves an average of only 13.9 clusters from cloud object storage when 
$n_\text{probe}$ is set to 240.
Furthermore, employing the more aggressive distance pruning threshold $\mathit{dist}_k^e$ further reduces the number of clusters, 
while the recall loss is 
less than $0.5\%$
at a baseline of $97\%$ recall. 

In addition, we exploit the RaBitQ distance estimators to perform dynamic pruning on active clusters (i.e., the clusters to be evaluated and not received yet) after dispatching the I/O requests. 
Specifically, while awaiting I/O responses from cloud object storage, data of distinct clusters inevitably arrives asynchronously over time. 
Therefore, the query engine can update the global ANN result set with completed requests, yielding 
a distance threshold which we set as the $k$-th distance of the vectors that have been evaluated so far and denote by
$dist_{k}^c$. 
This updated runtime threshold is then used to evaluate the remaining in-flight requests. 
If the $dist_{lo}$ for all vectors within a pending cluster exceeds the newly established $dist_{k}^c$, 
the query engine can safely skip that cluster.
Because this optimization is applied after network dispatching, it does not reduce the total number of issued requests, 
meaning the cost of cloud object storage remains unchanged.
Nevertheless, it improves end-to-end query throughput by dropping some network stragglers.

Our pruning strategy shares similar high-level goals with the query-aware dynamic pruning adopted in SPANN and SPFresh~\cite{spann,spfresh}.
However, our approach diverges fundamentally in its pruning criterion and mathematical foundations. 
SPANN and SPFresh rely on a ratio-based heuristic (i.e., discarding clusters whose distances to the query exceed 
that of the nearest cluster by a predefined multiplier), 
and provide no theoretical guarantees. 
Furthermore, this ratio introduces an additional hyperparameter that complicates user configuration and system setting. 
In contrast, our mechanism leverages the error bounds of the RaBitQ estimator to establish high-probability analytical error bounds, 
providing a theoretically grounded pruning criterion while eliminating the overhead of parameter tuning.

\begin{table}[t]
  \centering
  \small
  \caption{Number of remaining clusters after pruning. Each configuration is evaluated on an IVF index with $4,096$ clusters. The baseline recall for the No Prune configuration is omitted as pruning by $\mathit{dist}_k^u$ preserves nearly identical accuracy.}
  \label{tab:prune}
  \vspace{-2mm}
  \setlength{\tabcolsep}{3pt} 
  \begin{tabular}{l ccccc}
    \toprule
    & \multicolumn{2}{c}{\textbf{Prune by $\mathit{dist}_k^u$}} & \multicolumn{2}{c}{\textbf{Prune by $\mathit{dist}_k^e$}} & \textbf{No Prune} \\
    \cmidrule(lr){2-3} \cmidrule(lr){4-5}
    \textbf{Dataset} & \shortstack[c]{\textbf{Remaining} \\ \textbf{Clusters}} & \textbf{Recall} & \shortstack[c]{\textbf{Remaining} \\ \textbf{Clusters}} & \textbf{Recall} & \shortstack[c]{\textbf{Remaining} \\ \textbf{Clusters}} \\
    \midrule
    SIFT        & 60.0  & 0.9793 & 33.0  & 0.9781 & 80.0  \\
    MSong       & 16.9  & 0.9828 & 10.6  & 0.9819 & 30.0  \\
    Contriever  & 12.3  & 0.9751 & 6.2   & 0.9743 & 40.0  \\
    GIST        & 53.8  & 0.9705 & 23.9  & 0.9666 & 180.0 \\
    MXBAI       & 9.5   & 0.9756 & 5.7   & 0.9749 & 60.0  \\
    OpenAI      & 13.9  & 0.9716 & 8.1   & 0.9705 & 240.0 \\
    \bottomrule
  \end{tabular}
\vspace{-2mm}
\end{table}

\begin{figure}[tbp]
  \centering
  \includegraphics[width=0.9\linewidth]{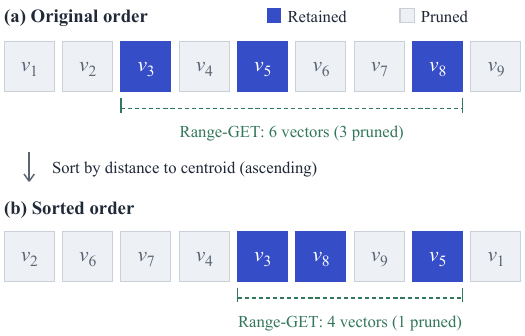}
  \Description{Range opt.}
  \caption{Illustration of range optimization.  With intra-cluster reordering, the un-pruned vectors become more concentrated within a cluster.}
  \label{fig:range}
\vspace{-2mm}
\end{figure}

\subsubsection{Range Optimization for Request Size Reduction}
\label{subsubsec:range}
Range optimization aims to selectively download only the precise byte range containing valid nearest-neighbor candidates within each target cluster for a given query.
Specifically, as detailed in Section~\ref{subsubsec:prune}, the system leverages the distance estimator of RaBitQ to generate pruning thresholds. 
With this pruning framework, the remaining un-pruned vectors typically occupy only a fractional slice of the cluster's total physical address space. 
As illustrated in Figure~\ref{fig:range}, for a cluster, our method dynamically issues a bounded range request covering the minimum contiguous byte 
sequence that encompasses all un-pruned candidates, which reduces overall download latency and conserves valuable network bandwidth.

As discussed in Section~\ref{subsubsec:reorder}, we reorder the vectors within a cluster by their distances to the cluster centroid.
Next, we provide more analysis on how this distance-based reordering facilitates the effectiveness of the aforementioned range optimization strategy.
RaBitQ~\cite{rabitq,exrabitq} transforms distance estimation between two vectors to an estimation of the inner product of two normalized vectors, as formalized below:
\begin{equation}
\begin{aligned}
    \|\mathbf{o}_r - \mathbf{q}_r\|^2 &= \|(\mathbf{o}_r - \mathbf{c}) - (\mathbf{q}_r - \mathbf{c})\|^2 \\
    &= \|\mathbf{o}_r - \mathbf{c}\|^2 + \|\mathbf{q}_r - \mathbf{c}\|^2 - 2\|\mathbf{o}_r - \mathbf{c}\| \cdot \|\mathbf{q}_r - \mathbf{c}\| \cdot \langle \mathbf{o}, \mathbf{q} \rangle
\end{aligned}
\label{eq:rabitq}
\end{equation}
where $\mathbf{o}_r$ and $\mathbf{q}_r$ represent the raw data and query vectors, respectively; 
$\mathbf{c}$ is the centroid vector used for normalization; 
and $\mathbf{o} := \frac{\mathbf{o}_r - \mathbf{c}}{\|\mathbf{o}_r - \mathbf{c}\|}$ and $\mathbf{q} := \frac{\mathbf{q}_r - \mathbf{c}}{\|\mathbf{q}_r - \mathbf{c}\|}$ 
denote the resulting normalized unit vectors. 
Within the IVF index, the vectors $\mathbf{o}_r$ within a cluster are normalized with the centroid of the cluster as the center $\mathbf{c}$.
During querying, the norms $\|\mathbf{o}_r - \mathbf{c}\|$ and $\|\mathbf{q}_r - \mathbf{c}\|$ are computed exactly, 
while RaBitQ provides an unbiased estimator for the inner product component $\left< \mathbf{o}, \mathbf{q} \right>$ subject to an analytical error bound.
Consequently, the estimation error of the squared Euclidean distance equals the estimation error of the inner product, scaled by a factor of $2\|\mathbf{o}_r - \mathbf{c}\| \cdot \|\mathbf{q}_r - \mathbf{c}\|$.
For a fixed query and cluster, $\| \mathbf{q}_r-\mathbf{c} \|$ remains static,
and a larger distance to the centroid ($\| \mathbf{o}_r-\mathbf{c} \|$) amplifies the inner-product estimation error.
Therefore, vectors farther from the centroid would have looser distance error bounds and are harder to prune.
This implies that vectors situated further from the cluster centroid inherently exhibit higher distance estimation error. 
Consider two data vectors within the same cluster that share a comparable true distance to the query vector. 
The vector positioned further from the cluster centroid would yield a looser distance lower bound ($dist_{lo}$), 
rendering it more difficult to be safely pruned. 
Consequently, sorting intra-cluster data vectors by their distance to the centroid effectively assembles these hard-to-prune vectors in the data layout. 
This spatial localization enables range optimization to further reduce byte volume retrieved from the cloud object storage.

{ Table~\ref{tab:range} evaluates the effectiveness of range optimization under different intra-cluster layouts.
Compared with disabling range optimization, enabling it reduces the volume of downloaded data by at least 44\%.
Intra-cluster reordering further increases this reduction.}
This consistent performance gain validates both our earlier analysis and the effectiveness of the intra-cluster reordering technique discussed in Section~\ref{subsubsec:reorder}.

\begin{table}[tbp]
  \centering
  \small
  \caption{Evaluating range optimization under varying data layouts at $97\%$ target recall. 
  For all datasets, the conservative threshold $\mathit{dist}_k^u$ is adopted for pruning. 
  The data volume downloaded is normalized by the volume without range optimization.}
  \label{tab:range}
  \vspace{-2mm}
  \begin{tabular}{lcc}
    \toprule
    & \multicolumn{2}{c}{\textbf{Data Volume Downloaded}} \\
    \cmidrule(lr){2-3}
    \textbf{Dataset} & \shortstack{\textbf{Random Vector}\\\textbf{Layout}} & \textbf{Reordered} \\
    \midrule
    SIFT        & 55.8\% & 48.8\% \\
    MSong       & 46.3\% & 34.1\% \\
    Contriever  & 41.5\% & 35.4\% \\
    GIST        & 30.7\% & 19.3\% \\
    MXBAI       & 34.6\% & 27.4\% \\
    OpenAI      & 22.2\% & 19.8\% \\
    \bottomrule
  \end{tabular}
\vspace{-2mm}
\end{table}

\subsubsection{Request Merging for Cost Reduction}
\label{subsubsec:merge}

\begin{figure}
  \centering
  \includegraphics[width=0.9\linewidth]{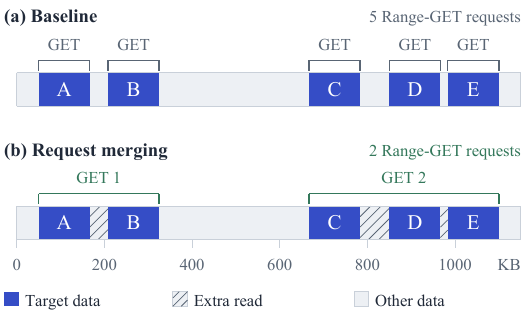}
  \Description{A toy example of request merging.}
  \caption{An illustrative example of request merging. The discrete Range-GET request count is curtailed by three after applying request merging with a 600~KB threshold.}
  \label{fig:merge}
\vspace{-2mm}
\end{figure}

After we apply the cluster pruning and range optimization techniques, 
we only need to fetch those byte ranges 
in the un-pruned clusters.
As introduced in Section~\ref{subsubsec:reorder},
instead of issuing a distinct I/O request for each 
un-pruned cluster,
which would incur high query costs,
we merge several I/O requests into a consolidated one
if the requests are closely situated within the index object.
We call this the \textit{request merging} strategy.

Following existing systems (e.g., Databricks~\cite{databricks} and Lance~\cite{lancedb}), we adopt a fixed threshold for  request merging.
Consequently, the size of any merged request is bounded by the threshold; requests whose individual size exceeds this threshold are naturally excluded from merging.
This is to prevent excessive payload sizes which may compromise end-to-end query throughput.
Based on empirical results reported in our experimental setup, we set the maximum merged-request size to 600~KB to balance per-request latency overhead and additional data transfer.
For latency-tolerant workloads, this threshold can be increased to further reduce the number of requests.
Figure~\ref{fig:merge} illustrates the retrieval of target byte ranges from the cloud object storage. 
Because the byte ranges of some requests are close, they can be retrieved via a single merged request. 

{ We design a dynamic programming (DP) algorithm to merge the requests under a fixed threshold \(T\), minimizing the request count and, for the same minimum count, the total downloaded bytes.
We first sort all candidate requests by start offset.
Let \(L(i,j)\) be the span covering requests \(i,\dots,j\), which form a feasible merged group if \(L(i,j)\le T\).
We define \(D[j]=(K_j,B_j)\) as the optimal result for the first \(j\) requests, recording the request count \(K_j\) and downloaded bytes \(B_j\).
Starting from \(D[0]=(0,0)\), we compute
\[
\begin{aligned}
D[j] = {}& \min_{\substack{1\le i\le j\\L(i,j)\le T}}
& \bigl(K_{i-1}+1,\;B_{i-1}+L(i,j)\bigr),
\end{aligned}
\]
where the comparison first minimizes request count, then downloaded bytes.
Each transition considers \(i,\dots,j\) as the last merged group.
We record the selected starting index and recover the grouping by backtracking from \(D[N]\).

We outline the proof of optimality below.
Requests larger than $T$ are excluded from the DP and added back as individual requests after merging.
Their fixed contribution to the request count and downloaded bytes does not affect optimality.
For the first remaining request, \(D[1]\) is clearly optimal.
Suppose that \(D[0],\dots,D[j-1]\) are optimal. 
Let an optimal strategy for the first \(j\) requests use requests \(i,\dots,j\) as its last merged group.
By the induction hypothesis, \(D[i-1]\) covers the preceding requests with no more requests and, if the request count is the same, no more downloaded bytes.
Since the DP considers every feasible starting index \(i\), it must find a solution at least as good as this optimal strategy.
Therefore, \(D[j]\) is also optimal.
By induction, \(D[N]\) minimizes the total number of requests and, among all consecutive groupings with that minimum count, the total downloaded bytes.}

\begin{table}[t]
  \centering
  \caption{Evaluation of request merging across different cluster ordering strategies. For each dataset, we build an IVF index with $4,096$ clusters and then measure the number of merged requests under a fixed number of scanned clusters ($97\%$ target recall).}
  \label{tab:merge}
  \vspace{-2mm}
  \small
  \setlength{\tabcolsep}{3pt}
  \begin{tabular}{lrrrrrr}
    \toprule
    & \multicolumn{6}{c}{\textbf{Requests After Merging}} \\
    \cmidrule(lr){2-7}
    \textbf{Dataset} & \shortstack{\textbf{Random}\\\textbf{Order}} & \shortstack{\textbf{Nearest}\\\textbf{Neighbor}} & \textbf{MST} & \textbf{Hilbert} & \textbf{Z-order} & \shortstack{\textbf{No}\\\textbf{Merge}} \\
    \midrule
    SIFT        & 60.3      & 25.5      & \textbf{24.6}  & 26.2  & 27.0      & 80.0 \\
    MSong       & 29.3      & \textbf{18.3}      & 19.0  & 23.3  & 23.6      & 30.0 \\
    Contriever  & 39.4      & \textbf{27.5}      & 28.1  & 34.8  & 35.0      & 40.0 \\
    GIST        & 175.9     & \textbf{150.1}     & 152.8 & 159.8 & 160.5     & 180.0 \\
    MXBAI       & 58.7      & \textbf{39.6}      & 40.7  & 50.5  & 50.6      & 60.0 \\
    OpenAI      & 236.9     & \textbf{218.4}     & 219.0 & 226.6 & 227.4     & 240.0 \\
    \bottomrule
  \end{tabular}
\vspace{-2mm}
\end{table}

Indeed, the spatial locality of clusters within the index object influences the effectiveness of request merging. 
If clusters are ordered randomly, the target clusters for a given query would be sparsely distributed across the index layout, leading to poor merging results. 
As established in Section~\ref{subsubsec:reorder}, we explore four different inter-cluster reordering strategies, discovering that the nearest-neighbor heuristic 
achieves the smallest
Hamiltonian path length across tested datasets. 
To validate this approach, we evaluate the end-to-end merging results for ANN search~\footnote{To isolate the empirical impact of request merging, cluster pruning and range optimization techniques are intentionally deactivated in this experiment, 
since the proposed request merging pipeline operates as a decoupled, modular framework which can be seamlessly integrated into alternative systems.}.
As illustrated in Table~\ref{tab:merge}, the nearest-neighbor heuristic 
achieves the smallest
total request count across most datasets. 
This empirically validates the use of Hamiltonian path length as a useful proxy for quantifying cluster spatial locality for downstream IVF-based ANN search.

\subsubsection{Local SSD Cache for Hot Clusters}
\label{subsubsec:cache}
When the compute node is configured with a local NVMe SSD,
it can serve as an intermediate cache for cloud object storage payloads 
~\footnote{Cloud vendors typically provide general-purpose block storage volumes (e.g., \textit{gp3} for AWS EC2~\cite{awsebs,awsec2}) attached to compute instances. 
However, these baseline volumes deliver limited default I/O throughput 
(e.g., capped at $3,000$ IOPS for \textit{gp3} if not paying extra fees), rendering them unsuitable for high-throughput updates.
Alternatively, users can provision specialized instances equipped with local, high-performance NVMe SSDs. 
Although these instances incur higher costs, for certain workloads the caching benefits can offset the overall system cost by reducing the API expenses of cloud object storage.}.
When an individual cluster is required during a query execution, the system first conducts a lookup in the local SSD cache; 
a cache hit completely bypasses the network downloading to the cloud object storage. 
We detail the cache of RemoteANN as follows.

The cache operates at cluster-level granularity mirroring the spatial access patterns executed during querying.
To reduce physical fragmentation within the cache, each cluster is padded to align with standard 4~KB pages. 
A single, pre-allocated backing file opened using the $\texttt{O\_DIRECT}$ flag (to bypass the OS's page cache) is used to cache all data.
Storage allocations within the backing file are managed at 4~KB granularity via a best-fit free-extent map that dynamically coalesces contiguous 
freed extents upon cluster eviction.
Following an empirical evaluation of various cache strategies, 
we use a caching engine combines TinyLFU-inspired admission~\cite{tinylfu} with aging LFU replacement, using per-cluster frequency counters
with periodic halving.
To improve I/O concurrency, we launch cache reads and writes asynchronously by leveraging the high-performance $\texttt{io\_uring}$ interface. 
Specifically, read operations are pooled, batched, and subsequently dispatched asynchronously; 
concurrently, cache insertion operations are completely decoupled from the critical query execution path and delegated to a background worker thread.

Note that there is an inherent architectural conflict between the cluster-level caching and our range optimization strategy of downloading. 
Specifically, the cache  manages data at cluster-level granularity, whereas range optimization aims to selectively retrieve minimal byte ranges of a cluster to conserve bandwidth. 
We resolve this tension by introducing an adaptive fallback threshold. 
When local NVMe caching is enabled, the range optimization is evaluated by a predefined threshold ratio $\gamma$ (e.g., $\gamma=0.8$). 
The system executes a range-optimized download request if and only if the optimized range spans less than $\gamma$ of the total cluster volume. 
For any request exceeding this ratio, the system downloads
the complete cluster, 
thereby trading marginal network overhead to populate the local SSD cache and benefit subsequent queries.
\begin{algorithm}[htbp]
\caption{Index Construction}
\label{alg:indexing}
\KwIn{Raw vector dataset $\mathcal{X}$, number of target clusters $K$}
\KwOut{RemoteANN index}

Use QGKMeans on $\mathcal{X}$ to obtain $K$ clusters (Section~\ref{subsubsec:symqg}) \\

Reorder the clusters and the vectors in each cluster (Section~\ref{subsubsec:reorder}) \\

Quantize each vector wrt the cluster centroid by RaBitQ and store the clustering results and prefixes of RaBitQ codes on the compute node and the suffixes of RaBitQ codes on cloud object storage, subject to the cluster-level ordering and vector-level ordering (Section~\ref{subsubsec:layout}) \\

\Return
\end{algorithm}

\begin{algorithm}[htbp]
\caption{Query Processing}
\SetKw{Continue}{continue}
\label{alg:querying}
\KwIn{Query vector $\mathbf{q}$, number of nearest neighbors $k$, number of target clusters $n_{\text{probe}}$}
\KwOut{ANN result}

Select the top $n_{\text{probe}}$ of clusters for $\mathbf{q}$ based on the IVF index \\
Estimate distances for all vectors in target clusters based on the 1-bit RaBitQ codes\\
Prune unpromising clusters for the query $\mathbf{q}$ (Section~\ref{subsubsec:prune})\\
For cache-hit clusters (un-pruned), retrieve data from local SSD (Section~\ref{subsubsec:cache}) \\
Perform range optimization for remaining clusters (Section~\ref{subsubsec:range}) \\
Perform request merging (Section~\ref{subsubsec:merge}) \\
Issue Range-GETs for merged requests to cloud object storage\\
Wait for I/O completion \hfill $\triangle$ Disk and cloud \\
Update cache \hfill $\triangle$ Background\\ 
Update ANN result based on the retrieved data\\
\Return ANN result
\end{algorithm}

\subsection{Summary}
\label{subsec:sum}

We summarize our indexing and querying process in Algorithm~\ref{alg:indexing} and Algorithm~\ref{alg:querying}, respectively.
Algorithm~\ref{alg:indexing} details the offline index construction process, 
which uses our efficient QGKMeans to accelerate IVF clustering and applies both intra-cluster and inter-cluster data
reordering to improve data locality.
Algorithm~\ref{alg:querying} presents how the query processing engine utilizes the data layout at runtime. 
By leveraging the local, memory-resident 1-bit RaBitQ codes, the query execution engine dynamically triggers 
cluster-level pruning, range optimization, request merging and caching. 
Together, these two parts formalize an algorithm-system co-design tailored specifically for the cloud object storage.
\section{Experiments}
\label{sec:exp}

In this section, we experimentally evaluate RemoteANN and baselines across various modern vector datasets with different dimensionalities and scales.
Moreover, we explore the effectiveness of our proposed strategies via an ablation study.

\subsection{Setup}
\label{subsec:setup}

\par\smallskip\noindent{\bf Datasets:} 
We employ six widely adopted million-scale datasets: SIFT, MSong, Contriever, GIST, MXBAI, and OpenAI. 
We also test the scalability of our method on larger datasets including Cohere and Deep-1B.
These datasets span diverse vector dimensionalities, cardinalities, and types and have been widely used in existing studies and benchmarks~\cite{annbench, bigann, douze2024faiss, superkmeans} to validate ANN algorithms and libraries. 
For datasets with a native evaluation partition, we utilize the default query workloads. 
For those lacking explicit query splits, we sample
$1,000$ vectors from the primary base dataset to serve as the evaluation query workload. 
The detailed statistics of these evaluation datasets are summarized in Table~\ref{tab:datasets}.

\begin{table}[t]
\centering
\small
\setlength{\tabcolsep}{4pt}
\caption{Statistics of datasets used for evaluation.}
\label{tab:datasets}
  \vspace{-2mm}
\begin{tabular}{llccc}
\toprule
\multicolumn{1}{l}{\textbf{Name}} & \multicolumn{1}{l}{\textbf{Data Type}} & \textbf{\# Vectors} & \textbf{\# Queries} & \textbf{Dim.} \\ 
\midrule
SIFT      & Image & 1,000,000   &  10,000   & 128   \\
MSong     & Audio & 992,272     &  1,000    & 420   \\
Contriever& Text  & 999,000     &  1,000    & 768  \\
GIST      & Image & 1,000,000   &  1,000    & 960   \\
MXBAI     & Text  & 769,382     &  1,000    & 1024  \\
OpenAI    & Text  & 999,000     &  1,000    & 1536  \\
Cohere    & Text  & 10,000,000  &  1,000    & 1024  \\
Deep-1B   & Image & 1,000,000,000 & 10,000    & 96   \\
\bottomrule
\end{tabular}
\vspace{-2mm}
\end{table}

{ 
\par\smallskip\noindent\textbf{Baselines:}
We evaluate RemoteANN against two representative baselines.

\noindent\textbf{Lance-RQ.}
Lance~\cite{lancedb} is an open-source storage engine with support for vector search over cloud object storage.
We use Lance v13.0.0 with its IVF-RQ index, denoted as Lance-RQ, which combines IVF partitioning with RaBitQ quantization.
To match RemoteANN's data placement, we keep the index metadata and RaBitQ codes (1-bit per dimension) in the EC2 instance's DRAM during querying,
while storing the full-precision vectors in cloud object storage for reranking.

\noindent\textbf{RemoteSPANN.}
We select SPANN~\cite{spann} as the second representative baseline since
its dynamic extension, SPFresh~\cite{spfresh}, is utilized by turbopuffer~\cite{turbopuffer},
a commercial vector database built specifically for object storage.
However, turbopuffer's implementation is closed-source, limiting our ability to instrument its execution and control its internal configuration.
In this study, we focus on static ANN workloads, so we build RemoteSPANN from the open-source SPANN implementation.
Specifically, we replace its local-disk I/O backend with the same cloud object storage layer used by RemoteANN,
enabling a controlled and fair comparison.
}

\par\smallskip\noindent{\bf System Setup:} 
Vector search contains two distinct phases: offline index construction and online query processing. 
A standard approach is to leverage a high-performance, resource-heavy machine to build the index, while provisioning cost-effective, lightweight compute nodes to handle online query serving. 
Following this model, the indexing runs on an on-premise server equipped with an Intel Xeon Gold 6418H@4.0GHz CPU (Sapphire Rapids architecture, 24 cores and 48 threads) and 
1~TB DRAM\footnote{Although the underlying physical layout is a dual-socket server containing two identical CPUs, cross-socket overheads degrade indexing 
efficiency for both our method and the evaluated baselines. Consequently, we employ NUMA hardware affinity tools ($\texttt{numactl}$) to strictly pin execution on a single CPU for all methods.}. 
We select Amazon Web Services (AWS) as our primary cloud evaluation platform.
Following index construction, the memory-resident routing structures are deployed on an Amazon EC2 compute instance, 
while the remaining data is persisted in an Amazon S3 bucket located within the same region as the compute node.
Unless stated otherwise, all experiments are executed on an AWS EC2 instance $\texttt{m5dn.2xlarge}$ (8 vCPUs, 32~GB DRAM, 300~GB local NVMe SSD, and up to 25~Gbps
network)\footnote{To enforce environment uniformity across the experiments, we deploy the $\texttt{m5dn.2xlarge}$ instance, as its local NVMe SSD allows us to validate the 
effectiveness of caching introduced in Section~\ref{subsubsec:cache}. Users can provision alternative instances lacking local NVMe SSD 
(e.g., $\texttt{c5n.2xlarge}$) for deployment that omits the cache layer. }. 
RemoteANN and RemoteSPANN are implemented in native C++, with cloud object storage I/O operations orchestrated by a download module built upon the official AWS SDK.
The offline indexing pipeline is compiled utilizing GCC 14.3.0 on an Ubuntu 22.04 LTS environment, whereas the online query execution engine is compiled via GCC 15.2.0 on the Ubuntu 26.04 LTS platform provided by the AWS EC2 instances.
For Lance-RQ, we use the official Python release which natively supports indexing, querying and remote I/O.

{

\par\smallskip\noindent\textbf{Parameter Settings:} 
During indexing, RemoteANN uses 9-bit RaBitQ quantization on all datasets, with the number of IVF clusters set according to the empirical recommendations of Faiss, as listed in Table~\ref{tab:kmeans}. 
For Lance-RQ, we use the same number of IVF clusters and retain the default values for all other indexing parameters.
For RemoteSPANN, we use the default indexing parameters of the open-source SPANN implementation~\cite{spann}.
During querying, we vary the number of probed clusters ($n_{\text{probe}}$) for RemoteANN and RemoteSPANN to evaluate their throughput--accuracy trade-offs.
For Lance-RQ, we follow the recommended tuning procedure by varying both $n_{\text{probe}}$ and \texttt{refine\_factor}, which controls the number of candidates reranked.
We consider \texttt{refine\_factor} values in $\{5, 10, 20\}$, extending this set to $\{5, 10, 20, 40, 80\}$ for Deep-1B, and report the throughput--accuracy Pareto frontier across the evaluated configurations.
}

\par\smallskip\noindent\textbf{Metrics for Query Performance:}
Following established indexing benchmarks~\cite{annbench,spann,bigann}, we evaluate query efficiency using throughput (quantified as queries per second (QPS)) and end-to-end latency, while search accuracy is measured via recall.
Across all experiments in this section, the search targets the retrieval of the top-$k$ nearest neighbors, with $k$ fixed at $10$.

\begin{figure*}[t]
  \centering
  \includegraphics[width=\linewidth]{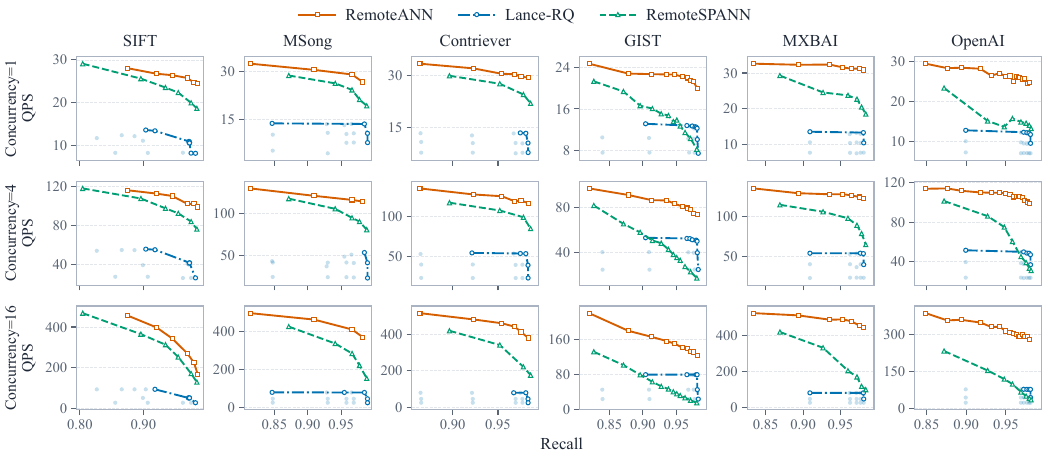}
  \Description{Main exp.}
  \caption{End-to-end search performance comparison across six million-scale datasets under varying query concurrency. 
    Each reported data point represents the mean throughput computed over three independent execution trials.}
  \label{fig:main-exp}
\end{figure*}

\subsection{Experimental Results}
\label{subsec:expres}

\subsubsection{Performance Study}
\label{subsubsec:perform}
This subsection presents end-to-end search performance, 
remote I/O statistics, 
index construction time, and memory footprint.

\par\smallskip\noindent{\bf End-to-end Query Performance:}
We systematically sweep the query-time parameters to evaluate throughput--recall trade-offs at concurrency levels of 1, 4, and 16. 
Figure~\ref{fig:main-exp} presents the resulting QPS--Recall curves, where curves closer to the top-right corner indicate better performance. 
Table~\ref{tab:latency_profile} reports the corresponding latency profiles at 97\% target recall~\footnote{For Lance-RQ, we report the latency at the configuration achieving the highest QPS among those reaching target recall.}.

We derive the following key observations.
(1) RemoteANN achieves better throughput--recall trade-offs than both Lance-RQ and RemoteSPANN across all evaluated datasets and concurrency levels. 
Its throughput advantage generally becomes more pronounced as concurrency increases, particularly on Contriever, MXBAI, and OpenAI.
(2) At 97\% target recall, RemoteANN achieves the lowest median ($p_{50}$) latency in all 18 configurations and the lowest tail ($p_{95}$) latency in 16 of them. 
Compared with Lance-RQ, its $p_{50}$ and $p_{95}$ latencies are lower by factors of $1.47$--$5.57$ and $1.19$--$5.15$, respectively. 
The only exceptions to its lowest tail latency occur on MSong at concurrency levels of 1 and 4, where RemoteSPANN has slightly lower $p_{95}$ latency.
(3) The QPS--Recall curves exhibit fluctuations, unlike the typically smoother curves of in-memory ANN indices~\cite{annbench,adsampling,gou2025symphonyqg,douze2024faiss}. These fluctuations are consistent with variability in cloud object storage I/O performance, as also reported in prior work~\cite{anyblob}.
(4) RemoteSPANN outperforms Lance-RQ on several datasets, demonstrating its competitiveness in these settings and supporting its inclusion as a complementary baseline.

\begin{table}[t]
  \centering
  \footnotesize
  \renewcommand{\arraystretch}{0.96}
  \caption{End-to-end query processing latency ($p_{50}$ and $p_{95}$, in milliseconds) at 97\% target recall under different concurrency levels (Conc.).}
  \label{tab:latency_profile}
  \vspace{-2mm}
  \setlength{\tabcolsep}{3pt}
  \resizebox{\columnwidth}{!}{%
  \begin{tabular}{lccccccc}
    \toprule
    & & \multicolumn{2}{c}{\textbf{RemoteANN}}
        & \multicolumn{2}{c}{\textbf{Lance-RQ}}
        & \multicolumn{2}{c}{\textbf{RemoteSPANN}} \\
    \cmidrule(lr){3-4}
    \cmidrule(lr){5-6}
    \cmidrule(lr){7-8}
    \textbf{Dataset} & \textbf{Conc.}
      & \textbf{$p_{50}$} & \textbf{$p_{95}$}
      & \textbf{$p_{50}$} & \textbf{$p_{95}$}
      & \textbf{$p_{50}$} & \textbf{$p_{95}$} \\
    \midrule
    \multirow{3}{*}{SIFT}
      & 1  & 35.6 & 72.9 & 86.8  & 111.3 & 40.8 & 81.9 \\
      & 4  & 36.9 & 74.4 & 91.0  & 119.7 & 41.3 & 84.3 \\
      & 16 & 62.6 & 98.4 & 314.1 & 343.0 & 89.7 & 125.0 \\
    \midrule
    \multirow{3}{*}{MSong}
      & 1  & 35.7 & 67.4 & 68.1  & 97.0  & 38.2 & 63.5 \\
      & 4  & 34.0 & 65.3 & 68.7  & 105.1 & 37.9 & 64.5 \\
      & 16 & 38.6 & 63.3 & 188.9 & 269.5 & 64.6 & 104.8 \\
    \midrule
    \multirow{3}{*}{Contriever}
      & 1  & 32.6 & 53.5 & 70.8  & 97.8  & 35.7 & 70.7 \\
      & 4  & 32.8 & 56.8 & 70.0  & 94.7  & 37.4 & 65.4 \\
      & 16 & 35.3 & 54.1 & 188.8 & 271.2 & 66.8 & 84.7 \\
    \midrule
    \multirow{3}{*}{GIST}
      & 1  & 49.4  & 84.2  & 72.8  & 102.1 & 93.8  & 141.9 \\
      & 4  & 49.2  & 83.5  & 72.8  & 99.2  & 168.3 & 257.7 \\
      & 16 & 118.5 & 221.4 & 189.1 & 272.9 & 703.0 & 803.8 \\
    \midrule
    \multirow{3}{*}{MXBAI}
      & 1  & 32.0 & 50.8 & 70.9  & 98.7  & 40.3 & 71.3 \\
      & 4  & 31.6 & 49.9 & 69.8  & 94.2  & 40.7 & 70.6 \\
      & 16 & 33.5 & 51.7 & 186.5 & 266.3 & 83.9 & 118.5 \\
    \midrule
    \multirow{3}{*}{OpenAI}
      & 1  & 37.3 & 68.3 & 76.0  & 103.4 & 66.7  & 106.8 \\
      & 4  & 37.6 & 64.9 & 76.6  & 97.9  & 87.8  & 250.3 \\
      & 16 & 44.3 & 70.4 & 191.3 & 280.4 & 281.3 & 397.6 \\
    \bottomrule
  \end{tabular}%
  }
\vspace{-2mm}
\end{table}
\begin{table}[t]
  \centering
  \footnotesize
  \setlength{\tabcolsep}{5pt}
  \caption{Average number of Range-GET requests per query to the cloud object storage at $97\%$ target recall.}
  \label{tab:reqnum}
  \vspace{-2mm}
  \resizebox{\columnwidth}{!}{%
    \begin{tabular}{lccc}
      \toprule
      & \multicolumn{3}{c}{\textbf{Number of Requests}} \\
      \cmidrule(lr){2-4}
      \textbf{Dataset} & \textbf{RemoteANN} & \textbf{Lance-RQ} & \textbf{RemoteSPANN} \\
      \midrule
      SIFT        & 18.5  & 97.1   & 58.5  \\
      MSong       & 10.7  & 57.6   & 39.9  \\
      Contriever  & 8.6   & 57.6   & 19.9  \\
      GIST        & 41.4  & 56.4   & 294.7 \\
      MXBAI       & 6.8   & 57.1   & 27.6  \\
      OpenAI      & 12.0  & 57.7   & 79.2  \\
      Cohere      & 18.6  & 26.2   & 300.0 \\
      Deep-1B     & 146.5 & 810.2  & --    \\
      \bottomrule
    \end{tabular}%
  }
\vspace{-2mm}
\end{table}

\par\smallskip\noindent{\bf Cloud I/O Statistics:} 
Table~\ref{tab:reqnum} reports the minimum average number of Range-GET requests per query achieved by each method under a recall constraint of at least 97\%.
As demonstrated, RemoteANN achieves the lowest number of distinct requests on all evaluated datasets, directly reducing the cost driven by the cloud object storage charging model.

\par\smallskip\noindent{\bf Indexing Time and Memory Footprint:} 
We evaluate the index construction latency and steady-state memory consumption\footnote{In this evaluation, 
we report in-memory construction time for RemoteANN and RemoteSPANN.
For Lance-RQ, we measure IVF training and shuffling, including their intermediate I/O but excluding the preliminary scan and final partition-index construction.
Concurrently, the memory footprint corresponds to the size required to maintain an active index. 
Under concurrent query workloads, memory consumption scales dynamically to accommodate buffers 
allocated for incoming cloud storage payloads.} across all compared methods and datasets, with the results summarized in Table~\ref{tab:indexinfo}.

\begin{table}[t]
  \centering
  \small
  \caption{Indexing time (T) and memory footprint (M).}
  \label{tab:indexinfo}
  \vspace{-2mm}
  \setlength{\tabcolsep}{3pt}
  \begin{tabular}{lcccccc}
    \toprule
    & \multicolumn{2}{c}{\textbf{RemoteANN}}
    & \multicolumn{2}{c}{\textbf{Lance-RQ}}
    & \multicolumn{2}{c}{\textbf{RemoteSPANN}} \\
    \cmidrule(lr){2-3}
    \cmidrule(lr){4-5}
    \cmidrule(lr){6-7}
    \textbf{Dataset}
    & \textbf{T (s)} & \textbf{M (GiB)}
    & \textbf{T (s)} & \textbf{M (GiB)}
    & \textbf{T (s)} & \textbf{M (GiB)} \\
    \midrule
    SIFT       & 9     & 0.03 & 7     & 0.04 & 116  & 0.18  \\
    MSong      & 15    & 0.08 & 7     & 0.09 & 224  & 0.52  \\
    Contriever & 20    & 0.12 & 12    & 0.13 & 628  & 0.93  \\
    GIST       & 27    & 0.15 & 15    & 0.16 & 596  & 1.17  \\
    MXBAI      & 23    & 0.13 & 13    & 0.14 & 461  & 0.96  \\
    OpenAI     & 43    & 0.23 & 23    & 0.25 & 1186 & 1.85  \\
    Cohere     & 252   & 1.72 & 180   & 1.91 & 7230 & 12.43 \\
    Deep-1B    & 11029 & 32.40 & 5176  & 30.62    & -    & -     \\
    \bottomrule
  \end{tabular}
\vspace{-2mm}
\end{table} 

As demonstrated, RemoteANN achieves up to a $31\times$ speedup in index construction over RemoteSPANN, aided by QGKMeans (Table~\ref{tab:kmeans}).
Although its construction takes longer than Lance-RQ, the times remain within the same order of magnitude.
This additional construction cost stems from two factors:
RemoteANN uses flat $k$-means rather than Lance-RQ's default hierarchical $k$-means~\footnote{Hierarchical $k$-means reduces training time but generally incurs higher clustering distortion~\cite{faissfaq}. Any resulting loss in recall may require probing more clusters to reach the same target recall, potentially increasing remote storage requests. We therefore use flat $k$-means to prioritize query performance, while QGKMeans reduces its construction cost.},
and it performs additional processes to improve query performance.
This additional construction cost is a one-time investment that pays off through the querying,
which we consider as a worthwhile trade-off for the substantial gains in query performance.
As for the memory consumption, RemoteANN and Lance-RQ are similar under the data-placement settings described in Section~\ref{subsec:setup}, 
whereas RemoteSPANN consumes substantially more memory. 
This difference arises primarily from their in-memory structures. 
Following SPANN~\cite{spann}, RemoteSPANN keeps cluster centroids and a routing graph in DRAM. 
Its bounded cluster sizes and vector replication result in a large number of clusters, 
with the centroid count reaching approximately 20\% of the dataset size in the default configuration. 
By comparison, memory usage in RemoteANN and Lance-RQ is dominated by the compact 1-bit RaBitQ codes, resulting in a much smaller footprint.

\begin{figure}
  \centering
  \includegraphics[width=0.9\linewidth]{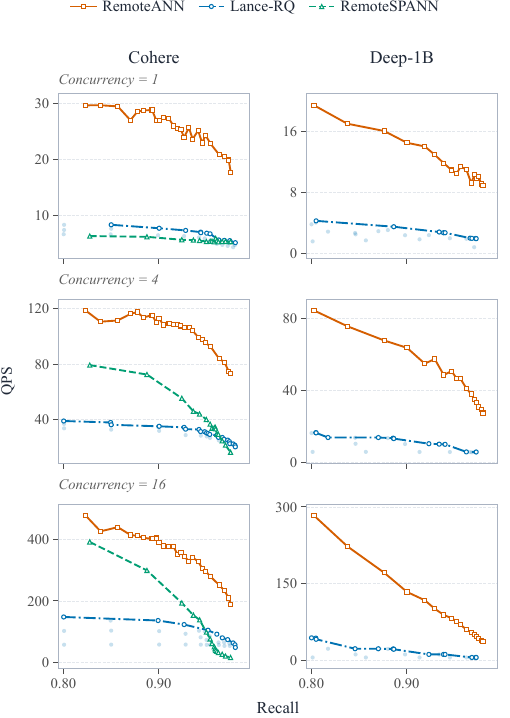}
  \Description{Scalability.}
  \caption{Evaluation of scalability.}
  \label{fig:scala}
\vspace{-2mm}
\end{figure}

\subsubsection{Scalability Evaluation}
\label{subsub:scala}
We validate the scalability of our method and baselines on Cohere and Deep-1B datasets. 
The corresponding index construction parameters (i.e., number of clusters in IVF) are detailed in Table~\ref{tab:kmeans}. 
For Deep-1B, we provision an AWS EC2 $\texttt{c5n.4xlarge}$ compute node to accommodate the elevated runtime memory footprint.
The end-to-end performance profiles are illustrated via the QPS-recall curves in Figure~\ref{fig:scala}, 
whereas the cloud object storage I/O metrics, indexing time, and memory footprint are summarized in Table~\ref{tab:reqnum} and Table~\ref{tab:indexinfo}. 
Results for RemoteSPANN are omitted on Deep-1B since the indexing process of SPANN exceeded the 24-hour time threshold.
Consistent with the empirical trends established on the million-scale datasets, 
RemoteANN consistently outperforms evaluated baselines on the Cohere and Deep-1B across all query concurrency levels. 
The stable search performance validates the scalability and robustness of our method.
Moreover, RemoteANN delivers substantial performance advantages in terms of the cloud I/O metrics,
directly mirroring the presented empirical results on the million-scale datasets.

\begin{figure}[tbp]
  \centering
  \includegraphics[width=0.9\linewidth]{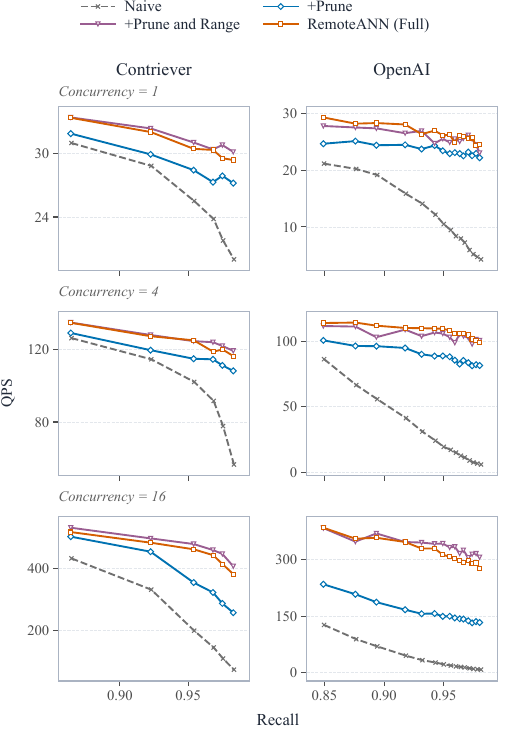}
  \Description{Ablation.}
  \caption{Ablation study of query optimization components.}
  \label{fig:ablation}
\vspace{-2mm}
\end{figure}

\begin{table}[tbp]
  \centering
  \small
  \caption{Cache hit ratio with 1 concurrent query at 97\% target recall across varying local cache capacities (normalized relative to the data volume in remote storage).}
  \label{tab:cache}
  \vspace{-2mm}
  \setlength{\tabcolsep}{8pt}
  \begin{tabular}{l ccc}
    \toprule
    & \multicolumn{3}{c}{\textbf{Ratio of Clusters from Cache (\%)}} \\
    \cmidrule(lr){2-4}
    \textbf{Dataset} & \textbf{10\% Cache} & \textbf{20\% Cache} & \textbf{30\% Cache} \\
    \midrule
    Contriever & 8.5\% & 14.6\% & 20.1\% \\
    OpenAI     & 7.3\% & 11.6\% & 13.1\% \\
    \bottomrule
  \end{tabular}
\vspace{-2mm}
\end{table}

\subsubsection{Ablation Study}
\label{subsubsec:ablation}
We evaluate the individual contributions of our query optimization components via an ablation study on the Contriever and OpenAI datasets. 
Specifically, we strip all query optimization strategies discussed in Section~\ref{subsec:query} to establish a naive baseline. 
We then incrementally integrate cluster pruning, range optimization and request merging to systematically reconstruct the complete RemoteANN framework. 
The resulting QPS-recall curves are presented in Figure~\ref{fig:ablation}. 
As demonstrated, both pruning and range optimization significantly elevate query throughput without compromising retrieval accuracy. 
Furthermore, incorporating the request merging mechanism introduces a minor throughput degradation under high-concurrency workloads due to the increased volume of downloaded data,
but it would effectively reduce cloud API operational costs (e.g., achieving 30\% reduction on Contriever and 14\% reduction on OpenAI according to Table~\ref{tab:prune} and Table~\ref{tab:reqnum}).

For the local NVMe SSD caching layer (detailed in Section~\ref{subsubsec:cache}),
during evaluations on the Contriever and OpenAI datasets, query throughput matched that of the cacheless baseline. 
Therefore, we omit the overlapping QPS-recall curves to isolate and analyze the cache hit ratio as the primary metric. 
As detailed in Table~\ref{tab:cache}, the cache hit ratio scales monotonically with expanded cache capacity. 
Furthermore, caching efficacy is heavily dictated by the underlying query workload characteristics, where the Contriever dataset yields a higher hit ratio compared to OpenAI.

Finally, we evaluate a variant of our framework, RemoteANN-Raw (denoted \textit{Raw} in Table~\ref{tab:raw}), 
which stores raw data vectors on cloud object storage instead of the compact RaBitQ quantization codes, 
while retaining all other query-side optimizations.
As presented in Table~\ref{tab:raw}, it achieves lower QPS than vanilla RemoteANN.
However, RemoteANN-Raw is still much more efficient than other baselines,
which indicates that the bulk of RemoteANN's speedup over the baselines stems mainly from our co-designed optimizations for indexing and querying, 
which remain effective regardless of whether the underlying payload is quantized or raw.

\begin{table}[t]
  \centering
  \small
  \caption{Query throughput comparison under varying concurrency levels (Conc.) at $97\%$ target recall.}
  \label{tab:raw}
  \vspace{-2mm}
  \setlength{\tabcolsep}{3pt}
  \resizebox{\columnwidth}{!}{%
  \begin{tabular}{llcccc}
    \toprule
    \textbf{Dataset} & \textbf{Conc.}
      & \textbf{RemoteANN} & \textbf{Raw}
      & \textbf{Lance-RQ} & \textbf{RemoteSPANN} \\
    \midrule
    \multirow{3}{*}{Contriever}
      & 1  & 29.5  & 27.5  & 13.3 & 24.4 \\
      & 4  & 120.0 & 109.5 & 54.2 & 99.0 \\
      & 16 & 412.5 & 305.2 & 78.7 & 221.1 \\
    \midrule
    \multirow{3}{*}{OpenAI}
      & 1  & 25.7  & 22.7  & 12.4 & 14.6 \\
      & 4  & 104.7 & 85.53  & 49.9 & 39.1 \\
      & 16 & 298.2 & 199.2 & 78.2 & 50.7 \\
    \bottomrule
  \end{tabular}%
  }
\vspace{-2mm}
\end{table}
\section{Related Work}
\label{sec:related}

\noindent\textbf{Indices for Approximate Nearest Neighbor Search.}
Existing index structures for Approximate Nearest Neighbor (ANN) search predominantly encompass 
graph-based~\cite{hnsw,nsg,2019diskann}, quantization-based~\cite{pq,pqfs,opq,addq,guo2020accelerating,rabitq,exrabitq}, 
hashing-based~\cite{gan2012locality, datar2004locality, huang2015query, indyk1998approximate, sun2014srs, tao2010efficient}, 
cluster-based~\cite{spann,spfresh}, and tree-based~\cite{beygelzimer2006cover, arya1993approximate, gu2022parallel, mtree} methods. 
While prior research has predominantly focused on in-memory or disk-based scenarios, 
RemoteANN is specifically designed for both indexing and querying over the cloud object storage.

\smallskip\noindent\textbf{Cloud Object Storage and Vector Search.}
Driven by the elasticity, durability, and cost-effectiveness of cloud object storage, both academia and industry have proposed different data management systems with an object store backend. 
AnyBlob~\cite{anyblob} explored the potential of utilizing the cloud object storage for OLAP workloads and developed a high-performance download manager.
A recent empirical study~\cite{li2025bench} benchmarked the performance of two representative disk-based ANN indices  (SPANN~\cite{spann} and DiskANN~\cite{2019diskann}) deployed over cloud object storage backends.
In industry, a series of vector search engines on cloud object storage has emerged, including Turbopuffer~\cite{turbopuffer}, Amazon S3 Vectors~\cite{s3vec}, Lance~\cite{lancedb}, and Databricks~\cite{databricks}. 
Both Turbopuffer and Amazon S3 Vectors adopt SPFresh~\cite{spfresh} as their underlying index structure, whereas Lance and Databricks utilize Inverted File (IVF) index variants.
We do not directly compare RemoteANN with the closed-source services turbopuffer and Amazon S3 Vectors, or with the full Databricks platform.
Instead, we evaluate Lance's IVF-RQ index as Lance-RQ, alongside RemoteSPANN, to compare representative indexing approaches under controlled storage and data-placement settings,
focusing on the algorithmic co-design of index structures and the cloud environment.

\smallskip\noindent\textbf{Accelerating IVF Construction.}
Recent work accelerates the $k$-means clustering in IVF construction by adopting techniques from ANN search.
SuperKMeans~\cite{superkmeans} combines dimension pruning~\cite{adsampling} with the PDX data layout~\cite{kuffo2025pdx} to reduce the cost of distance computation. 
Graph-based approaches also accelerate this step~\cite{lindormvector,scalakmeans,deng2018knn}.
For example, SeededANN~\cite{scalakmeans} uses an HNSW-based index over centroids and exploits assignments from previous iterations to seed subsequent searches. 
Another direction performs clustering on compressed vectors to reduce computation and memory overhead~\cite{kuffo2026stop}. 
Input compression represents an orthogonal design dimension to the choice of centroid-assignment method and could potentially be combined with our QGKMeans. 
Our evaluation focuses on acceleration using the original input vectors; 
we therefore leave combinations with input compression outside the scope of this comparison.
\section{Conclusion and Discussion}
\label{sec:conclusion}

In this paper, we presented RemoteANN, a vector search engine explicitly co-designed for cloud object storage. 
Motivated by the pronounced network latency and per-request pricing of the storage tier, 
RemoteANN couples an IVF index with RaBitQ quantization and a locality-aware data layout, 
accelerates index construction via QGKMeans, and introduces a query engine combining cluster pruning, range optimization, request merging, and caching. 
When evaluated on real-world datasets, RemoteANN consistently outperforms all representative baselines in query throughput and incurs lower storage API expenses.
Our ablation study further confirms the contribution of query-side optimizations, and our scalability evaluation demonstrates that these gains persist at billion-scale.

Several limitations point to directions for future work. 
Currently, RemoteANN targets a static IVF index.
Supporting efficient incremental updates and deletions remains an open challenge for vector search on cloud object storage. 
Moreover, our evaluation focuses on the warm-query setting, where index metadata is already resident in the compute node's memory. 
A deeper investigation of cold-start mitigation strategies, such as workload-driven prefetching, would further broaden RemoteANN's applicability to bursty or intermittent workloads. 
We believe the co-design principles introduced in this work offer a general foundation for future systems in this space.

\bibliographystyle{ACM-Reference-Format}
\bibliography{main}

\end{document}